\documentstyle[11pt]{article}
\newcommand{\bdis}{\begin{displaymath}}
\newcommand{\edis}{\end{displaymath}}
\newcommand{\beqn}{\begin{equation}}
\newcommand{\eeqn}{\end{equation}}
\newcommand{\beqna}{\begin{eqnarray}}
\newcommand{\eeqna}{\end{eqnarray}}
\newcommand{\nn}{\nonumber}

\newcommand{\bm}{\boldmath}
\newcommand{\epc}{\epsilon}

\newcommand{\del}{\partial}
\newcommand{\dcsb}{D$\chi$SB}
\def\AJP(#1,#2,#3){Aust. J. Phys. {\bf #1} (19#2) #3}
\def\NP(#1,#2,#3){Nucl. Phys. {\bf #1} (19#2) #3}
\def\NPBPS(#1,#2,#3){Nucl. Phys. {\bf B}(Proc. Suppl.){\bf #1} (19#2) #3}
\def\PL(#1,#2,#3){Phys. Lett. {\bf #1} (19#2) #3}
\def\PR(#1,#2,#3){Phys. Rev. {\bf #1} (19#2) #3}
\def\PRL(#1,#2,#3){Phys. Rev. Lett. {\bf #1} (19#2) #3}
\def\PTP(#1,#2,#3){Prog. Theor. Phys. {\bf #1} (19#2) #3}
\def\PTPS(#1,#2,#3){Prog. Theor. Phys. (Suppl.){\bf #1} (19#2) #3}

\newcommand{\DGL}{{\rm DGL}}
\newcommand{\QCD}{{\rm QCD}}
\newcommand{\eff}{{\rm eff}}
\newcommand{\conf}{{\rm conf}}
\newcommand{\qk}{{\rm quark}}
\newcommand{\qg}{{\mbox{\scriptsize\rm q-g}}}
\newcommand{\Yu}{{\rm Y}}
\newcommand{\Cou}{{\rm C}}
\newcommand{\Lag}{{\cal L}}
\newcommand{\tr}{{\rm tr}}
\newcommand{\Tr}{{\rm Tr}}
\newcommand{\Del}{{\cal D}}
\newcommand{\gpar}{\alpha_{\rm e}}
\newcommand{\Nf}{N_{\rm f}}
\newcommand{\Nc}{N_{\rm c}}
\newcommand{\Pc}{p_{\rm c}}
\newcommand{\simle}{\stackrel{<}{\scriptstyle\sim}}
\newcommand{\simla}{\stackrel{>}{\scriptstyle\sim}}
\begin{document}
\title{
Correlation between
Confinement and Chiral-Symmetry Breaking \\
in the Dual Higgs Theory
} 
\author{
S.~Umisedo,
H.~Suganuma and H.~Toki
\\
\em
Research Center for Nuclear Physics (RCNP), \\
Osaka University, Ibaraki, Osaka 567, Japan
}
\maketitle
\begin{abstract}
We study the correlation between confinement and dynamical chiral-symmetry 
breaking (D$\chi$SB)
in the dual Ginzburg-Landau (DGL) theory using the effective potential 
formalism. The DGL theory is an infrared effective theory based on the 
dual Higgs mechanism, and provides the nonperturbative gluon propagator,   
which leads to the linear quark potential.
The screening effect for the confining potential can be obtained 
by introducing the infrared cutoff corresponding to the hadron size.
We formulate the effective potential for D$\chi$SB in the DGL theory, 
and find the vacuum instability against quark condensation. 
To extract confinement effect, we separate the effective potential into
the confinement part and others by dividing the confinement term 
from other terms in the gluon propagator in the DGL theory.  
The confinement part provides the dominant contribution to D$\chi$SB,
which is regarded as monopole dominance for D$\chi$SB.  
The relevant energy for D$\chi$SB is found to be the infrared region 
below 1GeV.
Monopole dominance for the DGL propagator is also found in the intermediate
region, 0.2fm $\simle r \simle$ 1fm.
\end{abstract}


\section{Introduction}

Quark confinement and dynamical chiral-symmetry breaking(\dcsb) are the
most important nonperturbative phenomena in quantum chromodynamics(QCD)
	\cite{itzykson}.
Color confinement is characterized by the formation of
the squeezed color-electric flux tube
	\cite{haymaker}
and the linear potential between quark and anti-quark
	\cite{rothe,huang},
which confines the quarks inside hadrons.
On the other hand, \dcsb~is characterized by quark condensate,
large effective quark mass and the appearance of the
Nambu--Goldstone(NG) bosons as pions
	\cite{chengli}.
Although the QCD Lagrangian is chirally symmetric in the massless quark
limit, it is spontaneously broken in the nonperturbative vacuum
due to the quark condensate
$\langle \bar{q} q \rangle \neq 0$.
Such phenomena concerning with \dcsb~has been demonstrated and studied 
by using the effective models of QCD
	\cite{chengli,leutwyler,higashijima,miransky},
for instance Nambu--Jona-Lasinio model
	\cite{hk}.
These studies suggest that
light quarks get a large effective mass
	\cite{nachtmann},
$M \simeq $350MeV,
in the infrared energy region as a result of \dcsb.
The absence of the parity doubling in the
hadronic spectrum is understood in terms of \dcsb,
and the pion properties are understood by its identification to
the the NG boson.
As for confinement,
it is getting clear that quark confinement can be understood
in terms of the dual superconductor picture in the 't Hooft abelian gauge
	\cite{agf}
from recent studies of the lattice gauge theory
	\cite{kronfeld,hioki,kitahara}. 
In this picture, the QCD-monopole appearing by the abelian gauge fixing
condenses in the QCD vacuum,
and therefore quarks are confined through the dual Meissner effect
	\cite{nambu,thooft2,mandel}.

The close relation between these two nonperturbative phenomena
	\cite{banks}
have been 
suggested by the coincidence between the deconfinement phase transition and 
the chiral restoration in the lattice QCD
	\cite{kogut}.
Within the effective theory approach,
we have shown that QCD-monopole condensation plays an essential role on \dcsb
	\cite{como94,suganpb,ptp94,baryon95,ajp}.
Independently, 
Miyamura
	\cite{miyamura} 
and Woloshyn
	\cite{woloshyn}
found the similar result of {\em monopole dominance for D$\chi$SB}
in the lattice gauge theory.
The DGL theory is an infrared effective theory of the nonperturbative QCD,
where the QCD vacuum is described in terms of 
the abelian gauge theory with QCD-monopoles
	\cite{suzuki}.
In this theory,
it was shown that monopole condensation
induces large dynamical quark mass using the Schwinger-Dyson(SD) equation,
a non-linear integral equation for the quark mass function
	\cite{como94,suganpb,ptp94}.
The chiral phase transition at finite temperature was also studied
by solving the SD equation.
A strong correlation was found between the string tension
and the critical temperature of the chiral symmetry restoration
in terms of QCD-monopole condensation
	\cite{baryon95}.
Thus,
the above results suggest strong correlation between
confinement and \dcsb.

To get deeper insight on this correlation,
it is desirable to examine the contribution
of confinement to \dcsb ~more directly.
In the DGL theory, the nonperturbative gluon propagator is composed of
two parts, confinement term and others.
As for the inter quark potential,
the former leads to the linear potential and latter to the Yukawa potential.
To examine the contribution of each part, we calculate the effective potential
in the ladder level approximation, where
we can simply extract the contribution of confinement to \dcsb.
In Sec.2, we review the description of confinement
in the DGL theory.
We derive the nonperturbative gluon propagator from the DGL Lagrangian,
and see how it reproduces the $q$--$\bar{q}$ confining potential.
In Sec.3, we formulate the effective potential in terms of
the quark propagator.
The effective potential is formulated in the ladder level approximation
with the QCD renormalization improvement for the gauge coupling.
An analysis of the nonperturbative gluon propagator is done also.
In Sec.4, numerical results are discussed.
We also compare our results with the QCD-like theory,
where the perturbative gluon propagator is used in the ladder approximation.
Sec.5 is devoted to summary
and concluding remarks.

\section{Dual Higgs Mechanism and Quark Confinement}

It was pointed out that the quantum chromodynamics(QCD), a non-abelian
gauge theory, reduces to an abelian gauge theory containing monopoles 
in the 't Hooft abelian gauge
	\cite{agf},
which is defined by diagonalizing an appropriate gauge-dependent variable
$X[A_\mu(x)]$.
Since this gauge fixing condition leaves abelian gauge degrees of freedom unfixed,
the theory holds the local symmetry on the abelian gauge transformation.
In this gauge, only the diagonal components of the gluon behave as
abelian gauge fields and the remaining off diagonal components behave as 
charged matter fields under the residual abelian gauge transformation.
One of the most important features of the 't Hooft abelian gauge is that
monopoles appear at the singular points in gauge fixing condition
	\cite{agf}.

From the recent analyses based on the lattice QCD, 
the remaining off-diagonal gluon is largely reduced in  
the maximally abelian (MA) gauge
	\cite{magauge} 
and nonperturbative features are almost reproduced only by the abelian gluon 
including QCD-monopoles
	\cite{hioki,miyamura,woloshyn}.
For instance, in the SU(2) lattice gauge theory, MA gauge is defined 
by maximizing
\beqn
R	= \sum_{s,\hat{\mu}} {\rm Tr}[\sigma_3 U_\mu(s)
				      \sigma_3 U_\mu^\dagger(s)]
\eeqn
by the gauge transformation.
Here, $U_\mu(s) \equiv e^{iagA_\mu(s)}$ denotes the lattice link variable.
In this gauge, $X[A_\mu(x)]$ to be diagonalized is given by the Lorentz scalar,
\beqn
X(s)	= \sum_{\mu} [U_\mu(s) \sigma_3 U_\mu^\dagger(s)
		     +U_\mu^\dagger(s-\mu) \sigma_3 U_\mu(s-\mu)].
\eeqn
In the continuum limit, the MA gauge fixing is equivalent to 
$(\del^\mu \pm iaA_3^\mu)A_\mu^{\pm} = 0$,
where $A_3^\mu$ and $A^\mu_{\pm}$ denote the diagonal gluon and 
the off-diagonal gluons, respectively.
In the lattice formalism, 
the SU(2) link-variable in MA gauge is factorized as
$U_\mu=e^{i(\theta_\mu^+T^- + \theta_\mu^-T^+)}e^{i\theta_\mu^3T^3}$,
where the abelian link-variable $u_\mu \equiv e^{i\theta_\mu^3T^3}$ 
corresponds to the abelian part of $U_\mu$.
Similarly to the Wilson loop, 
the abelian Wilson loop, defined by the loop integral 
$w=e^{i\oint ds^\mu \theta_\mu^3T^3}$, 
obeys the area law and reproduces the string tension well in MA gauge
	\cite{hioki}.  
Therefore, the only abelian gluon seems relevant in MA gauge for 
confinement in the lattice QCD, which is called as abelian dominance 
for confinement.
Thus, for the description of the confinement physics, QCD in MA gauge 
would be well approximated by the abelian gauge theory including monopoles.

The abelian gauge field coupled to electric current $j_\mu$ and magnetic
current $k_\mu$ is described by introducing the dual gauge field $B_\mu$ as
	\cite{zwanziger}
\beqn
\Lag_{\rm Zwanziger}
	= K_{\rm gauge}(A,B) - j_\mu  A^\mu - k_\mu  B^\mu .
\eeqn
The kinetic term of gauge fields 
$ K_{\rm gauge}(A_\mu ,B_\mu )$ 
is expressed in the Zwanziger form
	\cite{suganpb,suzuki},
\beqn
 K_{\rm gauge} (A_\mu ,B_\mu ) 
	\equiv 
  - 
    \frac{2}{n^2}
    [n\cdot (\del \wedge A)]^\nu 
    [n\cdot ^*(\del \wedge B)]_\nu 
  - 
    \frac{1}{n^2}
    [n\cdot (\del \wedge A)]^2
  - 
    \frac{1}{n^2}
    [n\cdot (\del \wedge B)]^2, 
\label{eqn:KinGau}
\eeqn
which manifests the duality of the gauge theory. 
Here, $n_\mu$ is an arbitrary constant space-like 4-vector corresponding to
the Dirac string direction.
The diagonal gauge field
$A_\mu $ 
and the dual gauge field 
$B_\mu $ 
are defined on the Cartan sub-algebra 
$\vec H=(T_3,T_8)$: 
$A^\mu  \equiv A^\mu _3 T_3 + A^\mu _8 T_8$, 
$B^\mu  \equiv B^\mu _3 T_3 + B^\mu _8 T_8$.

The dual Ginzburg-Landau (DGL) theory is an infrared effective theory of QCD
based on the dual Higgs mechanism in the abelian gauge
	\cite{como94,suganpb,ptp94,suzuki,ptp120,conf95,ichie}.
Its Lagrangian is described by quarks $q$, monopoles $\chi$, and diagonal
part of gauge fields as
	\cite{ichie}
\beqn
    \Lag_\DGL 
	=
    \tr K_{\rm gauge}(A_\mu ,B_\mu )
  + \bar q (i\not\! \del - e \not\!\! A - m) q 
  + \tr [\Del_\mu , \chi ]^\dagger [\Del^\mu , \chi ]
  - \lambda \tr (\chi^\dagger \chi - v^2)^2, 
\label{eqn:DGLlag}
\eeqn
where 
$ \Del_\mu  \equiv  \del_\mu + igB_\mu $ 
is the dual covariant derivative with 
the dual gauge coupling 
$g$
obeying the Dirac condition,
$eg=4\pi $ 
	\cite{suganpb}.
The QCD-monopole field 
$\chi $ 
is defined on the nontrivial root vectors 
$E_\alpha $:
$\chi  \equiv \sqrt{2} \sum_{\alpha =1}^3 \chi _\alpha E_\alpha $,
with
$\sqrt{2}E_1 \equiv T_1 + iT_2$,
$\sqrt{2}E_2 \equiv T_4 - iT_5$,
$\sqrt{2}E_3 \equiv T_6 + iT_7$.
The QCD-monopoles condense in the vacuum, due to its self interaction.
Its condensation squeezes out the color electric flux 
and leads to linear inter-quark potential
	\cite{suganpb,ptp94,suzuki}.

Let us derive the nonperturbative gluon propagator
in the QCD-monopole condensed vacuum
	\cite{suganpb}. 
At the mean field level for the monopole condensate
$|\chi_\alpha|=v$,
one obtains the Lagrangian density as
\beqn
    \Lag_{\rm MF}
    	=
    \tr K_{\rm gauge}(A_\mu ,B_\mu )
  + \bar q (i\not\! \del - e \not\!\! A - m) q 
  + \frac{1}{2}m^2_B \vec{B}^2
\label{eqn:MFlag},
\eeqn
where 
$B_\mu $ 
acquires a mass of 
$m_B=\sqrt{3}gv$.
By integrating out the dual gauge field $B_\mu$, one obtains
the nonperturbative gluon propagator for $A_\mu$
\beqn
    D_{\mu \nu }(k)
	=
  - {1 \over k^2}
	\{
	  g_{\mu \nu }
       	+ (\gpar - 1){k_\mu k_\nu \over k^2} 
       	\}
  - {1 \over k^2} 
	{m_B^2 \over k^2 - m_B^2}
	{1 \over (n\cdot k)^2}
	\epc^\lambda  \, _{\mu \alpha \beta}
	\epc_{\lambda \nu \gamma \delta}
	n^\alpha n^\gamma k^\beta k^\delta ,
\eeqn
with the gauge parameter $\gpar$.
The double pole
$1/(n \cdot k)^2$
in the second term causes the infrared linear part
in the static quark-antiquark potential.
Hence, the second term is regarded as the
nonperturbative part of the gluon propagator.
The above expression, however, is derived without taking into
account the dynamics of light quarks.
With the dynamical quarks in the vacuum, the flux tube between
static charges when the distance exceeds some typical length
as hadron size, will be cut into two tubes with pair-produced
quarks on its end.
As a result, the static potential will be screened in the
QCD vacuum.
For the study of chiral symmetry, which is a symmetry 
of light quarks, we have to take into account the screening effect 
by the light quark polarization
	\cite{suganpb}.
In this case, the nonperturbative part is modified with the
infrared cutoff parameter 
$a$ 
corresponding to the inverse of hadron size as
	\cite{ptp94} 
\beqn
    D_{\mu \nu }(k)
	=
  - {1 \over k^2}
	\{
	  g_{\mu \nu }
       	+ (\gpar - 1){k_\mu k_\nu \over k^2} 
       	\}
  - {1 \over k^2} 
	{m_B^2 \over k^2 - m_B^2}
	{1 \over (n\cdot k)^2 + a^2}
	\epc^\lambda  \, _{\mu \alpha \beta}
	\epc_{\lambda \nu \gamma \delta}
	n^\alpha n^\gamma k^\beta k^\delta .
\label{eqn:DDGL}
\eeqn

Now we derive the static inter-quark potential.
Introducing external color-electric source $\vec{j}$,
the current action can be obtained as
\beqna
Z_j   & = &\int {\cal D}A e^{i \int d^4x \Lag_{\rm MF}(A,j)}
	=  \int {\cal D}A \exp\left\{i \int d^4x[{1 \over 2}
					\vec{A}^\mu(x) D^{-1}_{\mu \nu}(x)
					\vec{A}^\nu(x)
				      -
				        \vec{j}_\mu(x) \vec{A}^\mu(x)
				        ]\right\}		\nn	\\
      & = &\exp\left\{i \int d^4x \int d^4y [-{1 \over 2}
			\vec{j}^\mu(x) D_{\mu \nu}(x-y) \vec{j}^\nu(y)
			]\right\}
	\equiv  e^{-i V[j] \int dt}.
\label{eqn:Gfcnl}
\eeqna
For the static quark and 
anti-quark current
\beqn
\vec{j}^\mu (x)
	=  \vec{Q}g^{\mu 0}[\delta^3 (\mbox{\bm $x$} - \mbox{\bm $b$})
			    -
			    \delta^3 (\mbox{\bm $x$} - \mbox{\bm $a$})],
\eeqn
whose Fourier component is given as
\beqn
\vec{j}^\mu (k)
	=  \vec{Q}g^{\mu 0}2\pi\delta(k^0)
			   (e^{-i \mbox{\scriptsize\bf k} \cdot 
			   	  \mbox{\scriptsize\bf b}}
			   -e^{-i \mbox{\scriptsize\bf k} \cdot 
			   	  \mbox{\scriptsize\bf a}}),
\label{eqn:current}
\eeqn
the current action
$S_j$
is expressed as
\beqna
	S_{j}
 & = & -\frac{1}{2}
        \int \!\! d^4xd^4y \,
          \vec{j}^\mu(x)\!\cdot\! 
          D_{\mu \nu}(x-y) 
          \vec{j}^\nu(y) 					\nn \\
 & = &  \frac{1}{2}
        \int \!\! \frac{d^4k}{(2\pi)^4}\,
          \vec{j}^\mu(-k)\!\cdot\!
          \left[
                \frac{1}{k^2-m^2_B}g_{\mu \nu} 
	      + \left\{
	      		\frac{m^2_B - a^2}{k^2 - m^2_B}
	      	      + \frac{a^2}{k^2}
	      	\right\}
	      	\frac{1}{(n \cdot k)^2 + a^2}g_{\mu \nu}
          \right] 
          \vec{j}^\nu(k)					\nn \\
 & = & -\vec{Q}^2 \int dt \int \frac{d^3\mbox{\bm$k$}}{(2\pi)^3}
	( 1 - \cos\mbox{\bm $k$}\cdot\mbox{\bm $r$} )
 	\left[\frac{1}{\mbox{\bm $k$}^2 + m^2_B}
	     +\frac{ m^2_B ( \mbox{\bm $k$}^2 + a^2 ) }
	           { \mbox{\bm $k$}^2 ( \mbox{\bm $k$}^2 + m_B^2 ) }
 	      \frac{1}{(\mbox{\bm $n$} \cdot \mbox{\bm $k$})^2 + a^2}
 	\right].
\label{eqn:Sj}
\eeqna
Here, we have used 
$n^\mu = (0,\mbox{\bm $n$})$
for the Dirac string direction with unit vector {\bm $n$}
and the relative coordinate
{\bm $r$} = {\bm $b$} $-$ {\bm $a$}.
The first term produces the short-range Yukawa potential
while the second term gives rise to the infrared linear potential.

We set 
$\mbox{\bm $n$} /\!\!/ \mbox{\bm $r$}$
from the symmetry of the system 
and the energy minimum condition.
Similarly in the argument for the Abrikosov vortex,
there appears the physical ultraviolet cutoff corresponding to
the coherent length in the integration of the transverse momentum
$k_{\rm T}$
	\cite{suganpb},
due to the reduction of the monopole condensate in the core region
of the flux tube.
Quantitatively, the inverse of the coherent length is given by the
monopole mass $m_\chi = 2\sqrt{\lambda}v$, and the dual gluon mass
$m_B$ vanishes in the core region
$r_{\rm T} \simle m_\chi^{-1}$,
which leads to the cutoff on the ultraviolet region
$k_{\rm T} \simla m_\chi$
in the  above integral.
Thus, the static potential is obtained as
\beqn
V(r)	= - \frac{\vec{Q}^2}{4\pi}
	    \frac{e^{- m_B r}}{r}
	  + \sigma \frac{1 - e^{-ar}}{a},
\label{eqn:Vr}
\eeqn
with string tension given as
\beqn
\sigma	= \frac{\vec{Q}^2}{8\pi}
	  \left\{
	        (m^2_B - a^2)\ln \left(
	        		      \frac{m^2_\chi + m^2_B - a^2}
	        		           {m^2_B - a^2}
	        		\right)
	        +    a^2    \ln \left(
	             		      \frac{m^2_\chi - a^2}{a^2}
	             		\right)
	  \right\}.
\label{eqn:stringtension}
\eeqn
The second term of Eq.(\ref{eqn:Vr})
reduces to the linear potential $\sigma r$
in the limit of 
$a \rightarrow 0$,
which corresponds to the absence of the screening effects.
In the region $r \simla a^{-1}$, screening effect suppress the
potential and it approaches to a constant value in the limit 
$r \rightarrow \infty $.
In Fig.1, we show the static potential in the DGL theory
for various infrared screening cutoff $a$
together with the phenomenological Cornell potential
	\cite{lucha},
\beqn
V_{\rm Cornell}(r)
	= - {e^2_\Cou \over 3\pi} \cdot {1 \over r}
	  + k_\Cou r  ,
\eeqn
with
$ e_\Cou \simeq 2$
and
$ k_\Cou \simeq 1$GeV/fm.
The DGL parameter set is to be chosen so as to reproduce the
string tension
$\sigma = 1$GeV/fm,
and the flux-tube radius of hadrons as
$m^{-1}_B = 0.3 \sim 0.4$ fm.
Here, we set $m_B = 0.5$GeV,
$e = 5.5$,
and choose
$m_\chi =$1.26, 1.17 and 1.06 GeV
for $a =$ 0, 85, 150 MeV respectively.
In the recent lattice QCD simulation
	\cite{latticescreen},
the screening effect by dynamical quarks is observed
at the infrared region as $r \simla $2fm,
which corresponds to $a \simle$ 100 MeV.

There are two possibilities of modification on the
short range part
	\cite{conf95}.
One is the vanishing of the dual gauge mass $m_B$
and the other is the running of the gauge coupling.
Though the Yukawa type part is obtained by integrating
the first term of the Eq.(\ref{eqn:Sj}) without cutoff,
$m_B$ should vanish in the ultraviolet region as noted above.
This may cause the shift of the Yukawa type potential
to a Coulomb type one in the short range.
Also, in this region, the gauge coupling constant is expected
to be modified.
In the DGL theory, the off diagonal gluons are omitted
on the basis of abelian dominance.
However, this feature is expected only for the long range
physics and in the perturbative region all components of
the gluon become relevant.
Hence, with the increase of the number of coupling gluons, 
the description only with abelian components becomes insufficient.
To compensate the contribution of off diagonal components,
the coupling constant should be modified in this region.
Considering these modifications one may find better coincidence
between the static potential presented by DGL theory and
the phenomenological one in the short rage.

In any case, the dual Ginzburg--Landau theory describes the
global feature of the confinement dynamics well.
In addition, as stated in the introduction, this model exhibits 
\dcsb~for quarks as well.
Therefore one can inquire much about these nonperturbative
physics through this model, instead of dealing with the
highly complex full QCD in the infrared region.

\section{Correlation between \dcsb~and Confinement}

Recently, the essential role of QCD-monopole condensation 
on D$\chi $SB has been pointed out
by solving the Schwinger-Dyson (SD) equation 
	\cite{como94,suganpb,ptp94}. 
In this chapter, we study the correlation between \dcsb~and confinement
from the view point of energy density, {\em i.e.} in the effective
potential formalism.
Approximating the full quark propagator as 
$iS^{-1}= \not\!\!\, p - M(p^2) + i\epc $, 
one obtains the SD equation for the quark mass function 
$M(p^2)$
in the Euclidean space,
\beqn
    M(p^2)
	= 
    \int\!\!{d^4q \over (2\pi )^4} 
    \vec Q^2 {M(q^2) \over q^2 + M^2(q^2)} 
    D_{\mu \mu}(q-p)
\label{eqn:SDE}
\eeqn
in the chiral limit.
Here,
$\vec{Q}^2
	=
	{\Nc - 1  \over  2\Nc}e^2
$
denotes the abelian electric charge of the quark.
In this equation,
$D_{\mu \mu }(k)$ 
is composed of three parts, 
\beqn
    D_{\mu \mu}(k)
	= 
   {2m_B^2 \over k^2(k^2 + m_B^2)} \cdot
   {n^2k^2 + a^2 \over (n \cdot k)^2 + a^2}
  +{2 \over k^2 + m_B^2}
  +{1 + \gpar \over k^2}
	=
    D_{\mu \mu}^\conf(k)
  + D_{\mu \mu}^\Yu(k)
  + D_{\mu \mu}^\Cou(k).
\label{eqn:GpropTr}
\eeqn
The first term 
$D_{\mu \mu }^\conf(k)$ 
is responsible for the linear confinement potential 
	\cite{como94,suganpb} 
in the absence of the screening effect,
$a=0$. 
The second term 
$D_{\mu \mu }^\Yu(k)$ 
is related to the short-range Yukawa potential,
and the Coulomb part 
$D_{\mu \mu }^\Cou(k)$ 
does not contribute to the quark static potential. 
In spite of this decomposition,
it is difficult to separate and compare each contribution 
to \dcsb~in the nonlinear SD equation.
To examine these contributions separately,
it is useful to study the effective potential
which corresponds to the vacuum energy density.
In this paper, we study the effective potential
	\cite{cjt}
variationally in the chiral limit.

Within the ladder level approximation, 
the effective potential 
$V_\eff[S]$ 
leads to the ladder SD equation when imposed the extremum condition 
in terms of the full quark propagator 
$S(p)$ 
	\cite{higashijima}.
Using the nonperturbative gluon propagator 
$D^{\mu \nu}(k)$ 
in the DGL theory, the effective potential as a functional of
the dynamical quark mass 
$M(p^2)$ 
is expressed as 
\beqn
    V_\eff[S]
    	=
    i\Tr \ln[S_0 S^{-1}]
  + i\Tr [SS_0^{-1}]
  + \int\!\! {d^4p \over (2\pi)^4}{d^4q \over (2\pi)^4}
   {\vec Q^2 \over 2}
    \tr[\gamma _\mu  S(p) \gamma _\nu  S(q) D^{\mu \nu}(p-q) ],
\label{eqn:EffAct}
\eeqn
where 
$S_0 (p)$ 
is the bare quark propagator, 
$iS_0^{-1}(p) = \not\!\!\, p - m + i\epc $ 
. 

Here, we would like to comment on the wavefunction renormalization.
Considering the effect of the wavefunction renormalization, the ladder
SD equation for the quark propagator,
$iS^{-1}(p)=z^{-1}(p^2)(\not\! p - M(p^2) ) $,
is given by coupled equations
\beqna
	M(p^2)
	  & = &
     	z(p^2)
     	\int\!\! {d^4q \over (2\pi)^4i} \vec{Q}^2 z(q^2)
     	\frac{M(q^2) g^{\mu \nu}}
     	     {q^2 - M^2(q^2)}
     	D_{\mu \nu},		\label{eqn:sigma}	\\
     	z^{-1}(p^2)
     	  & = &
     	1
     -	{1 \over p^2}
     	\int\!\! {d^4q \over (2\pi)^4i} \vec{Q}^2 z(q^2)
     	\frac{p^\mu q^\nu + p^\nu q^\mu - p \!\cdot\! q\,g^{\mu \nu}}
     	     {q^2 - M^2(q^2)}
     	D_{\mu \nu}.		\label{eqn:z}
\eeqna
Suppose one uses the perturbative gluon propagator and 
the Higashijima--Miransky approximation 
	\cite{higashijima,miransky} 
for the running coupling constant
\beqn
	e^2(k^2)
	\simeq
	e^2(\bar{k}^2),
\eeqn
where 
$k=p-q$, $\bar{k}^2 \equiv \max(p^2,q^2)$.
Then, Eq.(\ref{eqn:z}) reduces to the form
\beqn
	z^{-1}(p^2)
	  =
	1
     +  \gpar{2 \over p^2}
     	\int\!\! {d^4q \over (2\pi)^4i} \vec{Q}^2(\bar{k}^2) 
     	\frac{z(q^2)}
     	     {q^2 - M^2(q^2)}
     	{p\cdot q  \over (p-q)^2},
\eeqn
which leads
$z(p^2) \equiv 1$
in the Landau gauge
($\gpar = 0$).
In this case, Eq.(\ref{eqn:sigma}) reduces to Eq.(\ref{eqn:SDE}).
On the other hand, with the use of the nonperturbative gluon propagator in 
Eq.(\ref{eqn:DDGL}), this is not necessarily the case.
In the Landau gauge, however, we found
$z(p^2) \simeq 1$
even in the DGL theory by solving the coupled SD equations
including $z(p^2)$
	\cite{como96}.

Hence, working in the Landau gauge,
we approximate the wavefunction renormalization as
$z(p^2) \simeq 1$,
which corresponds to the neglect of the
second term in the right hand side of Eq.(\ref{eqn:z}).

In the effective potential formalism, its 2-loop term is given as
\beqna
	V^{(2)}
	&=& 
	2\Nf\Nc\!\!
	\int\!\!\!{d^4q \over (2\pi)^4i}\!\!
	\int\!\!\!{d^4p \over (2\pi)^4i}
	\vec{Q}^2 z(q^2) z(p^2)
	\frac{M(q^2)M(p^2)g^{\mu \nu}
	    + p^\mu q^\nu  + p^\nu q^\mu
	    - p \!\cdot\! q \,g^{\mu \nu}
	      }
	     {[M^2(q^2) - q^2]
	      [M^2(p^2) - p^2]
	      }
	D_{\mu \nu}					\nn\\		
	&=&
	2\Nf\Nc\!\!
	\int\!\!\!{d^4q \over (2\pi)^4i}\!\!
	\int\!\!\!{d^4p \over (2\pi)^4i}
	\vec{Q}^2 z(q^2) z(p^2)
	\frac{M(q^2)M(p^2)}
	     {[M^2(q^2) - q^2][M^2(p^2) - p^2]}
	D^\mu{}_{\mu}					\nn\\
	&+&
	2\Nf\Nc\!\!
	\int\!\!\!{d^4p \over (2\pi)^4i}
	\frac{p^2}
	     {M^2(p^2) - p^2}
	\{1-z(p^2)\},
\label{eqn:V2true}
\eeqna
where Eq.(\ref{eqn:z}) is used.
Here, to set the effect of the wavefunction renormalization as
$z(p^2) \equiv 1$ is equivalent to
drop off the second term of Eq.(\ref{eqn:V2true}).
This approximation would hold good in the Landau gauge
because the coupled SD equations shows $z(p^2) \equiv 1$ 
	\cite{como96}.

Hence, the effective potential consistent with the approximation
$z(p^2) \equiv 1$ is given by 
\beqna
    V_\eff[M(p^2)]
	&\equiv&
    V_\qk[M(p^2)] + V_\qg[M(p^2)]		\nn \\
    	&=&
  - 2\Nf \Nc 
    \int\!\! {d^4p \over (2\pi)^4}
    \{
      \ln({p^2 + M^2(p^2) \over p^2})
     -2{M^2(p^2) \over p^2 + M^2(p^2)}
    \}						\nn \\
	&-&
    \Nf(\Nc - 1) 
    \int\!\! {d^4p \over (2\pi)^4} {d^4q \over (2\pi)^4}
    e^2 {M(p^2) \over p^2 + M^2(p^2)} 
   {M(q^2) \over q^2 + M^2(q^2)} 
    D_{\mu \mu}(p-q)  \label{eqn:EffPot}  	  
\eeqna
in the Euclidean space.
It is easy to check that the extremum condition on Eq.(\ref{eqn:EffPot})
with respect to $M(p^2)$ leads to the SD equation (\ref{eqn:SDE}) 
	\cite{suganpb,ptp94}
exactly.

In Eq.(\ref{eqn:EffPot}),
the first term $V_\qk$ is the one-loop contribution as in Fig.2a.
The second term $V_\qg$ with 
$D_{\mu \mu }$ 
is the two-loop contribution with the quark-gluon interaction
as expressed in Fig.2b.
The important point is that this second term
$V_\qg$
is divided into three parts as
\beqn
	V_\qg
	=
	V_\conf + V_\Yu + V_\Cou ,
\eeqn
corresponding to the decomposition of 
$D_{\mu \mu }$ 
in Eq.(\ref{eqn:GpropTr}).
Hence, by estimating
$V_\conf$, $V_\Yu$, $V_\Cou$
respectively, it is possible to examine each contribution to \dcsb.

Before going further, remember that 
the nonperturbative gluon propagator depends on the Dirac string direction
$n_\mu$.
However, due to the
$q$-$\bar{q}$
pair polarization effect around the quark,
the Dirac-string direction 
$n_\mu $
becomes indefinite
	\cite{ptp94}.
Therefore we take the average value on
$n_\mu $,
\beqn
	\langle
	{n_\mu n_\nu \over
	 (n \cdot k)^2 + a^2 }
	\rangle _{\rm av}
   =
   	{1 \over 2\pi^2}
   	\int \!\!\! d\Omega_n
   	{n_\mu n_\nu \over
	 (n \cdot k)^2 + a^2 }
   =
   	{k_\mu k_\nu \over k^2}f_{/\!\!/}(k^2)
   +	\left\{ \delta_{\mu \nu}
   		- {k_\mu k_\nu \over k^2}
   	\right\}f_{\perp}(k^2),
\eeqn
with
\beqna
	f_{/\!\!/}(k^2)
  &=&
   	{1 \over (a + \sqrt{a^2 + k^2} )^2 },		\\
	f_{\perp}(k^2)
  &=&
  	\frac{ a + 2\sqrt{a^2 + k^2} }
  	     { 3a (a + \sqrt{a^2 + k^2} )^2 }
   =	{1 \over 3}
   	\left\{
   		{2(a^2+k^2) \over ak^2(a + \sqrt{a^2 + k^2})}
   		-
   		{1 \over k^2} 
   	\right\},
\eeqna   	
where the angle integration is performed in the Euclidean
4-dimensional space.
We then obtain the averaged gluon propagator 
$\bar{D}_{\mu \nu}$ as
\beqn
	\bar{D}_{\mu \nu}
   =	
   	{1 \over 3}
   	\left\{	\delta_{\mu \nu}
   		-
   		{k_\mu k_\nu \over k^2}
   	\right\}
   	\left\{	{1 \over k^2}
   		+
   		{2 \over k^2 + m_B^2}
   		+
   		{4m_B^2 \over k^2(k^2 + m_B^2)}
   		{a^2 + k^2 \over a(a + \sqrt{a^2 + k^2})}
   	\right\}
   +	
   	\gpar{k_\mu k_\nu \over (k^2)^2}
\label{eqn:barDmunu}
\eeqn
which requires the modification to Eq.(\ref{eqn:GpropTr})
only on confinement part as
\beqn
	D^\conf_{\mu \mu}
	\rightarrow
	\frac{4m_B^2}
	     {k^2 (k^2 + m_B^2)}
	\frac{k^2 + a^2}
	     { a (a + \sqrt{k^2 + a^2} ) }.
\eeqn

Therefore the effective potential we ought to compute is given as the sum of
following four terms
\beqna
	V_\qk \;\,
 & = &  \hspace{10pt}
 	\frac{2N_{\rm f}N_{\rm c}}{(4\pi)^{2}} 
        \int^\infty_0 dp^{2}\! 
        \left\{ 
             - p^{2}\ln \left( 1+\frac{M^2(p^2)}{p^{2}} \right) 
             + \frac{2\,p^{2}M^2(p^2)}{M^2(p^2)+p^{2}} 
        \right\} 				\label{eqn:Vqk}	 \\
	V_{\rm C} \;
 & = &- \frac{2N_{\rm f}N_{\rm c}}{(4\pi)^{4}} 
        \int^\infty_0 dp^{2}\! 
        \frac{p^{2}M(p^2)}{M^2(p^2)+p^{2}} 
        \!\!\int^\infty_0 dq^{2}\! 
        \frac{q^{2}M(q^2)}{M^2(q^2)+q^{2}} 
        \vec{Q}^2
        \frac{1}{\max(p^2,q^2)} 				 \\
	V_{\rm Y} \:
 & = &- \frac{2N_{\rm f}N_{\rm c}}{(4\pi)^{4}} 
        \int^\infty_0 dp^{2}\! 
        \frac{p^{2}M(p^2)}{M^2(p^2)+p^{2}} 
        \!\!\int^\infty_0 dq^{2}\! 
        \frac{q^{2}M(q^2)}{M^2(q^2)+q^{2}} 
        \vec{Q}^2						\nn \\
 &   &  \hspace{18pt} \times
        \frac{4}{\pi} \!\!\int^\pi_0 d\theta \sin^{2}\theta \,
        \frac{1}{(p-q)^{2}+m^{2}_{B}} 				 \\
	V_\conf \;
 & = &- \frac{2N_{\rm f}N_{\rm c}}{(4\pi)^{4}} 
        \int^\infty_0 dp^{2}\! 
        \frac{p^{2}M(p^2)}{M^2(p^2)+p^{2}} 
        \!\!\int^\infty_0 dq^{2}\! 
        \frac{q^{2}M(q^2)}{M^2(q^2)+q^{2}} 
        \vec{Q}^2			 			\nn \\
 &   &  \hspace{18pt} \times
        \frac{8}{\pi a} \!\!\int^\pi_0 d\theta \sin^{2}\theta \,
        \frac{1}{a+\sqrt{(p-q)^{2}+a^{2}}} 
        \frac{m_{B}^{2}((p-q)^{2}+a^{2})}
             {(p-q)^{2} ((p-q)^{2}+m_{B}^{2})}.			
\eeqna
As for the running coupling, we adopt hybrid type one
	\cite{suganpb,ptp94}
in the Higashijima--Miransky approximation
	\cite{higashijima,miransky},
\beqn
	e^2((p-q)^2)
	=
	{48\pi ^2 (\Nc + 1) \over 11\Nc - 2\Nf}
	\left[\ln {\Pc^2 + \max(p^2,q^2) \over \Lambda^2_\QCD}
	  \right]
	  ^{-1}.
\eeqn
Choosing $\Pc$ as
\beqn
\Pc^2	= \Lambda_\QCD^2
	  \exp [ {48\pi^2 \over e^2} {\Nc + 1 \over 11\Nc - 2\Nf}],
\eeqn
the above expression connects smoothly the perturbative running coupling 
of QCD in the ultraviolet region and the infrared effective coupling $e$
of the DGL theory
	\cite{suganpb,ptp94}.
From the renormalization group analysis of QCD 
	\cite{higashijima}, 
the approximate form of the quark-mass function 
$M(p^2)$ 
is expected as 
\beqn
    M(p^2)
    	= 
    M(0) {\Pc^2 \over (\Pc^2 + p^2)} 
    \left[
       \ln \frac{\Pc^2}{\Lambda^2_\QCD}
       \left/
       \ln \frac{\Pc^2 + p^2}{\Lambda^2_\QCD}
       \right.
    \right]^
       {1 - {\Nc^2 - 1 \over 2\Nc} \cdot {9 \over 11\Nc - 2\Nf}}.
\label{eqn:MassAns}
\eeqn
Since the exact solution 
$M_{\rm SD}(p^2)$ 
of the SD equation (\ref{eqn:SDE}) 
	\cite{ptp94} 
is well reproduced by this ansatz (\ref{eqn:MassAns})
with 
$M(0) \simeq 0.4$GeV 
and 
$\Pc^2 \simeq 10\Lambda _\QCD^2$,
we use this form 
as a variational function of the effective potential.

At the end of this section, let us examine the nonperturbative gluon
propagator in the DGL theory.
By performing the Fourier transformation on Eq.(\ref{eqn:barDmunu}),
we show in Fig.3
the Coulomb, Yukawa and confinement parts of the DGL gluon propagator
in the coordinate space,
\beqna
D_{\mu\mu}^\Cou(x-y)
	&\equiv& 	\int \frac{d^4k}{(2\pi)^4}
		     \frac{1}{k^2}
		     e^{ik(x-y)}
	 = \frac{1}{4\pi^2}\frac{1}{(x-y)^2}				\\
D_{\mu\mu}^\Yu(x-y)
	&\equiv& 	\int \frac{d^4k}{(2\pi)^4}
		     \frac{2}{k^2 + m_B^2}
		     e^{ik(x-y)}					\\
D_{\mu\mu}^\conf(x-y)
	&\equiv&	\int \frac{d^4k}{(2\pi)^4}
		     \frac{4m_B^2}{k^2(k^2 + m_B^2)}
		     \frac{a^2 + k^2}{a(a + \sqrt{a^2+k^2})}
		     e^{ik(x-y)}
\eeqna
as the function of $r \equiv |x-y|$.
The confinement part gives a dominant contribution to the DGL propagator
in the whole region, especially 0.2fm $\simle r \simle$ 1fm.
This means that a strong correlation is brought by the confinement part
of the nonperturbative gluon propagator.
Since this confinement part is a direct consequence of the monopole
condensation, this significant contribution suggests 
{\em monopole dominance for the DGL propagator},
at least in the intermediate region.

\section{Numerical Results and Discussions}

We show in Fig.4 the effective potential
$V_\eff$
as a function of the infrared effective quark mass
$M(0)$,
using the mass function (\ref{eqn:MassAns}) with
$\Pc^2 = 10\Lambda^2_\QCD$.
We have used the same parameters as in Ref.
	\cite{suganpb,ptp94},
$\lambda =25$, $v=0.126{\rm GeV}$, $e=5.5$ and $a=85$MeV
so as to reproduce the inter-quark potential and 
the flux-tube radius 
$R \simeq 0.4 {\rm fm}$ 
	\cite{suganpb}.
It takes a minimum at finite
$M(0) \simeq $0.4GeV,
which means that the nontrivial solution is more stable
than the trivial one in terms of the energetical argument.
Hence, chiral symmetry is spontaneously broken.

We show in Fig.5
$V_\qk, V_\conf, V_\Yu$ and $V_\Cou$
as the function of $M(0)$.
The lowering of the effective potential contributes to \dcsb.
Although both the Yukawa and the Coulomb terms contribute slightly to
lower the effective potential, it is mainly lowered by the 
confinement part
$V_\conf$
and there is a large cancellation between $V_\qk$ and $V_\conf$.
Thus, \dcsb~is brought by 
$V_\conf$
arising from monopole condensation in the DGL theory
which is regarded as  monopole dominance for \dcsb.
Such a dominant role of the confinement effect on \dcsb~is found
for any value of $M(0)$.

In Fig.6, we show also the integrands of each term
\beqn
	V_\eff
	=
	\int^\infty_0 \!\!\!\!\!\!
	dp^2 v_\eff(p^2)
	=
	\int^\infty_0 \!\!\!\!\!\!
	dp^2 \{v_\qk(p^2)
     +  v_\conf(p^2)
     +  v_\Yu(p^2)
     +  v_\Cou(p^2)\}
\eeqn
to examine which energy region is important to \dcsb.
Here we have used the exact solution 
$M_{\rm SD}(p^2)$
of the SD equation (\ref{eqn:SDE})
	\cite{suganpb,ptp94} 
as the mass function to get rid of the ambiguity 
arising from the choice of the trial mass function.
Among $v_\conf$, $v_\Yu$ and $v_\Cou$ 
the confinement part $v_\conf$ is always dominant for all momentum region.
All the three terms, $v_\conf,v_\Yu,$ and $v_\Cou$,
contribute to the effective potential mainly in the low momentum region 
less than 1GeV although there are also long tails running into
high momentum region over 1GeV.
These behaviors directly reflect the
profile of the quark mass function $M_{\rm SD}(p^2)$.
It is notable that such contributions from high momentum region are 
strongly canceled in $v_\eff$ as shown in Fig.6.
Consequently, the remaining contribution only from low momentum region
($p^2 < 0.4 {\rm GeV}^2$)
plays an important role to \dcsb.
This seems consistent with the above result that infrared confinement
effect is dominant for \dcsb.

Now, we compare our results with those in the QCD-like theory
	\cite{higashijima,miransky,Italy,AKM},
which is a more familiar framework for \dcsb.
In the QCD-like theory, the perturbative gluon propagator 
without the confinement force is used and the effective potential
$V_\eff$ can be divided into two parts, $V_\qk$ and $V_\qg$
within the improved ladder approximation
	\cite{higashijima,miransky}.
Here, $V_\qk$ is given by Eq.(\ref{eqn:Vqk}),
and $V_\qg$ takes a simple form
	\cite{higashijima},
\beqn
	V_\qg
	=
	\int^\infty_0 \!\!\!\!\!\!
	dp^2 v_\qg(p^2)
	=
      - \frac{2N_{\rm f}N_{\rm c}C}{(4\pi)^{4}} \!\!\!
        \int^\infty_0\!\!\!\!\!\!\!\! dp^{2}\! 
        \frac{p^{2}M(p^2)}{M^2(p^2)+p^{2}} 
        \!\!\int^\infty_0\!\!\!\!\!\!\!\! dq^{2}\! 
        \frac{q^{2}M(q^2)}{M^2(q^2)+q^{2}} 
        \frac{3e^2(\max(p^2,q^2))}{\max(p^2,q^2)} 				 
\eeqn
with $C = {\Nc ^2 -1 \over 2\Nc}$ and the running coupling constant 
\beqn
	e^2(p^2)
	=
	{48\pi ^2 \over 11\Nc - 2\Nf}
	\{\ln {\Pc^2 + p^2 \over \Lambda^2_\QCD}
	  \}
	  ^{-1}.
\eeqn
The effective potential as a function of the infrared quark mass 
in the QCD-like theory also takes the double-well structure
when the gauge coupling constant is enough large
	\cite{higashijima}.
Similarly in the DGL theory, the driving force of \dcsb~is brought
by the quark-gluon interaction part $V_\qg$, and a large cancellation
between $V_\qk$ and $V_\qg$ is found there.

As for the relevant energy scale for \dcsb, we show in Fig.7
the integrands $v_\qk(p^2)$ and $v_\qg(p^2)$ using the solution
of the SD equation in the QCD-like theory with parameters,
$e(0) = 11$ and $\Lambda_\QCD = 746$MeV, which reproduce
$f_\pi = 93$MeV and $\langle \bar{q} q \rangle_{\rm RGI}=-(239$MeV$)^3$
	\cite{AKM}.
Again the relevant energy scale is found to be less than 0.5 GeV
for \dcsb, which suggests the importance of the treatment on the
infrared strong-coupling region, where large confinement effects
would appear.
In the DGL theory, the confining force has been included as a result of
monopole condensation, which has been supported in the
lattice gauge theory
	\cite{kronfeld,hioki,kitahara},
and therefore the DGL approach seems to provide a more consistent
picture for the infrared region in QCD.

In the QCD-like theory, large values for $e(0)$ and
$\Lambda_\QCD$ are often used to reproduce $f_\pi$ or the constituent
quark mass
	\cite{Italy,AKM}.
As an interesting possibility, such an ``enhancement" of the strong 
coupling region may be regarded as a compensation for the neglected
confinement effect in the QCD-like theory
	\cite{suganpb}.
In any case, it would be meaningful to include some confinement effect
in the QCD-like theory to argue whether the confinement effect is
relevant to \dcsb~in QCD
	\cite{Will}.

\section{Summary and Concluding Remarks}

We have studied confinement and dynamical chiral-symmetry
breaking (\dcsb) in the dual Ginzburg--Landau (DGL) theory 
using the effective potential formalism.
The DGL theory describes confinement by squeezing of the color-electric
flux through dual Higgs mechanism.
The resulting gluon propagator is composed of two parts,
one is the usual perturbative part and the other nonperturbative
term leading to the linear potential between static $q$--$\bar{q}$ system.
To investigate the correlation between \dcsb~and confinement
we have formulated the effective potential for the quark propagator.
Making use of the nonperturbative gluon propagator in the DGL theory
we have included the effect of confinement into the effective potential.
Within the ladder approximation with the renormalization group
improvement for the coupling,
the effective potential is 
formulated as a function of the dynamical quark mass
$M(p^2)$.

The effective potential has been calculated as a function
of the infrared quark mass 
$M(0)$
with the variational function suggested by renormalization group
analysis of QCD.
We have found the double-well structure of the effective potential,
so that the nontrivial solution is more stable
than the trivial one and leads to \dcsb.
To examine the role of confinement, the interaction term
$V_\qg$
has been divided into the confinement part
$V_\conf$
and others
($V_\Yu$,$V_\Cou$).
We have found that the confinement part
$V_\conf$
stemming from monopole condensation gives the dominant contribution to 
$V_\qg$
($V_\qg \simeq V_\conf$),
which means the {\em monopole dominance} for \dcsb.
It has been also found that the low momentum contribution from less than
1 GeV plays an important role for \dcsb.

Finally, we would like to consider the critical scale   
between the nonperturbative QCD (NP-QCD) and the perturbative QCD (P-QCD)
in terms of the dual Higgs mechanism
	\cite{ichiequlen97}.
In this framework, these regions are divided 
by the typical energy scale of QCD-monopole condensation, 
$m_B(m_\chi) \sim 1$GeV.
Above the energy scale, the system can be characterized by
the asymptotic freedom in P-QCD with the Coulomb interaction.
On the other hand, the infrared strong-coupling region  
would be characterized by monopole condensation in the dual Higgs theory, 
because QCD-monopoles seem to play relevant role to the 
nonperturbative phenomena such as confinement and \dcsb. 
It is interesting to compare this critical scale with a 
{\it normalization point} $\mu$ of the operator product expansion 
(OPE) in the QCD sum rule \cite{narison}, where the quantum fluctuations 
with the momentum scale above $\mu$ are included
in the Wilson coefficient, while those with the momentum scale 
below $\mu$ are included in the local operator.
Since the short-scale physics are approximately described by
the P-QCD while the long-scale physics are connected to nonperturbative 
phenomena, $\mu$ physically corresponds to the border of NP-QCD and P-QCD.
Hence, the scale of QCD-monopole condensation, $m_B(m_\chi) \sim$ 1GeV,
may provide the physical image of $\mu$ in OPE. 
In any case, we conjecture that QCD in the 't Hooft abelian gauge 
exhibits the two different theoretical aspects: 
one is the short-distance physics subject to P-QCD formulated 
in the {it trivial vacuum}, and the other is the infrared nonperturbative 
physics described by the dual Higgs theory with {it QCD-monopole condensation}.

\section*{Acknowledgment}
We are thankful to Dr. S.~Sasaki for the 
numerical program of the Schwinger--Dyson 
equation and fruitful discussions.

\section*{Figure Captions}
Fig.1: Static $q$--$\bar{q}$ potential in the DGL theory.
Phenomenological potential (dashed curve) is also shown for comparison.
With the increase of the screening cutoff $a$, the screening effect
becomes strong. The saturation of the potential is observed at the long
distance, $r \simla 1/a$.
The phenomenological potential is also shown for comparison.

Fig.2: The diagrams which contribute to the effective potential
in the ladder level approximation.
	(a) The quark loop contribution $V_\qk$ without the explicit interaction.
	(b) The two-loop diagram $V_\qg$ including the quark-gluon interaction.
Here, the curly line with a black dot denotes the nonperturbative gluon
propagator in the DGL theory.

Fig.3: The DGL gluon propagator in the coordinate space,
$D_{\mu\mu}(r) \equiv F.T. D_{\mu\mu}(k)$,
with $m_B = 0.5$GeV and $a = 85$MeV.
The linear part is significantly large compared to the other parts 
especially in the intermediate region 0.2fm $\simle r \simle$ 1fm.
Thus, the long-range strong interaction is mainly brought by
the confinement part in the DGL theory.

Fig.4: The total effective potential $V_\eff$ as a function of 
the infrared quark mass $M(0)$.
The nontrivial minimum appears at $M(0)\sim 0.4$GeV, which indicates
dynamical breaking of chiral symmetry.

Fig.5: $V_\qk$, $V_\conf$, $V_\Yu$ and $V_\Cou$ are is shown as a function
of $M(0)$. The confinement part $V_\conf$, 
plays the dominant role through the lowering the effective potential.

Fig.6: Integrands $v_\qk$, $v_\conf$, $v_\Yu$ and $v_\Cou$
of effective potential are shown as functions of the Euclidean momentum $p^2$. 
The confinement part $v_\conf$ is more significant than $v_\Yu$ and $v_\Cou$
for all momentum region.

Fig.7: The integrands $v_\eff$, $v_\qk$ and $v_\qg$ in the QCD-like theory.
These quantities are plotted as functions of the Euclidean momentum squared
$p^2$.
\end{document}